\documentclass[12pt,a4paper]{article}
\usepackage[T2A]{fontenc}
\usepackage[cp1251]{inputenc}
\usepackage[english]{babel}
\usepackage{amssymb}
\usepackage{amsmath}
\usepackage{graphicx}
\usepackage[small,sc,center]{titlesec}
\usepackage{indentfirst}
\usepackage[labelsep=period]{caption}
\usepackage{caption}
\newcommand{\msun}{\,$M_{\odot}$}

\newcommand{\ergcms}{\,erg\,cm$^{-3}$\,s$^{-1}$}
\newcommand{\ergs}{\,erg\,s$^{-1}$}
\newcommand{\gcmq}{\,g\,cm$^{-3}$}

\newcommand{\kms}{\,km\,s$^{-1}$}

\newcommand{\cmq}{\,cm$^{-3}$}
\newcommand{\cmsqg}{\,cm$^2$\,g$^{-1}$}
\newcommand{\cmmsq}{\,cm$^{-2}$}
\newcommand{\ha}{H$\alpha$}
\newcommand{\mgii}{Mg\,{\sc ii}}
\newcommand{\feii}{Fe\,{\sc ii}}
\newcommand{\oi}{O\,{\sc i}}
\newcommand{\oiii}{O\,{\sc iii}}
\newcommand{\oii}{O\,{\sc ii}}
\newcommand{\siii}{Si\,{\sc ii}}
\begin{document}
	
\begin{center}
\textbf{\large Superluminous supernova SN~2018ibb: Circumstellar shell and spectral effects}

	\vskip 5mm
\copyright\quad
2024 \quad N. N. Chugai \footnote{email: nchugai@inasan.ru}\\
\textit{ Institute of astronomy, Russian Academy of Science, Moscow} \\
Submitted  02.09.2024 
\end{center}

{\em keywords:\/} stars --- supernovae; stars --- supermassive stars

\noindent
{\em PACS codes:\/} 

\clearpage
 
 \begin{abstract}

I explore observational effects of the cirsumstellar gas around superluminous supernova 
SN~2018ibb.
High velocity \feii\ narrow absorptions are reproduced in the model of fragmented cold dense 
shell. 
Unusual selective absorption in the emission doublet of [\oi] is explained as an effect 
of the radiation scattering in \siii\ doublet lines in the supernova envelope.
The strong emission of [\oiii] doublet at $t_{max} + 565$ days is proposed to originate 
 from the supernova envelope, whereas its asymmetry is explained by the dust formation 
 in the unshocked ejecta.
Circumstellar interaction modeling combined with observational data suggests 
  the cirsumstellar shell mass of  $\sim 0.14$\msun. 
 
\end{abstract}

\section{Introduction}

Superluminous supernova (SLSN) SN~2018ibb at the redshift $z = 0.166$ draws a special  
  attention due to unique observational data and important consequences  for the 
 theory of supermassive stars and their explosion (Schulze et al. 2024).
 Authors conclude that in this particular case we see the pair-instability supernova (PISN)
 with the explosion energy of $\gtrsim10^{52}$\, erg and 25--44\msun\ of ejected $^{56}$Ni.
Spectra show absorption lines of \mgii\ 2796, 2803\AA\ arising in the circumstellar matter 
 (CSM) with the expansion velocity of 2918\kms. 
The CSM mass is not determined, yet it should be emphasized that the ejection of  
  massive shell ($\gtrsim 1$\msun) with the energy $\gtrsim 10^{50}$\,erg before the 
  final explosion PISN could raise a problem, since, in theory, an explosion of a supermassive star 
  with the large energy and the amount of ejected $^{56}$Ni should be a single event without 
  preceding large pulsation mass ejection (Woosley et al. 2002; Heger \& Woosley 2002).

The crucial role of the CSM mass for the theory of supermassive stars makes 
 the study of CSM around presupernova SN~2018ibb highly demanding.
Constraints on the mass and radius of a CS shell could be obtained via the SN/CSM shock   
  interaction modeling combined with relevant observational data. 
In this context one cannot help noticing three problems of SN~2018ibb spectra 
  interpretation that closely related to the CSM issue.
  
The first is a 20-years-old problem of high-velocity narrow absorption 
 lines of the multiplet 42 \feii\ (4924, 5018, 5169\,\AA) in SLSN spectra around the 
 light maximum. 
Kasen (2004) proposes that narrow \feii\ lines in SLSN SN~1999as form in the cold dense  
  shell (CDS) that is an outcome of SN/CSM interaction.
Yet he pointed out that a drawback of this interpretation is too small CDS width.
Alternatively, Moriya et al. (2019) explain narrow \feii\ lines in SLSN 
 SN~2007bi and SN~1999as introducing cut-off of ejecta density distribution at velocities 
  $\gtrsim 12000$\kms\ and neglecting the absorption by the CDS. 
 
The second problem is an unusual selective absorption in the  
 [\oi] 6300, 6364\,\AA\ emission doublet apparent at $t_{max}+ 287$ days.
This feature was attributed to [\oi] lines absorption in the CDS (Schulze et al. 2024).
However, the detailed estimate shows that this mechanism requires too large oxygen mass 
($\gtrsim 30$\msun) in the CDS. 
A conceivable alternative (explored below) might be a radiation scattering in 
 \siii\ 6347, 6371\,\AA\ doublet inside  undisturbed ejecta.

Finally, the problem of the line-emitting site for [\oii] and [\oiii] emission lines at 
  the nebular stage.  
Generally, these lines might be emitted by the CSM photoionized by the shock wave radiation 
  (Schulze et al. 2024). 
An alternative unexplored possibility is the emission of these lines by  undisturbed supernova ejecta.   
  
The present communication explores these problems of spectra interpretation, which 
  matter for the CSM issue.
The inferred results turn out to be useful constraints for the CSM model. 
In the next section I model 
  the CDS effect for \feii\ lines, the effect of \siii\ doublet in the [\oi] emission, 
  and calculate [\oiii] doublet for both locations of the line-emitting site.
In the third section I model the SN/CSM interaction and infer CSM mass.
In two Appendices I clarify some points related to narrow \feii\ lines and \siii\ 
excitation.  
  
\begin{figure}
	\centering
	\includegraphics[trim=0 200 0 100, width=\columnwidth]{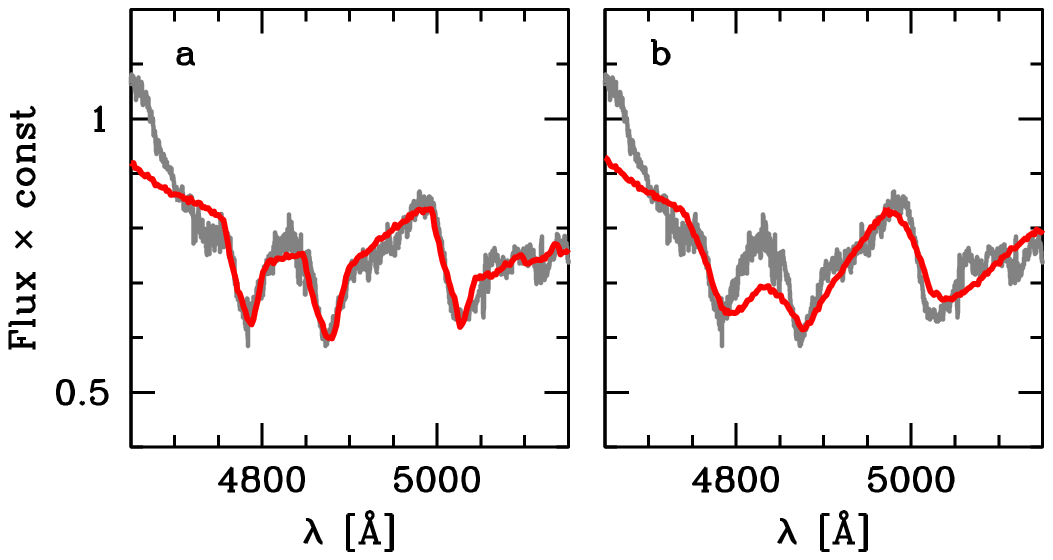}
	\caption{
  \feii\ lines for two model options ({\em red} line) compared to the observational spectrum 
  at $t_{max} + 33$ days (Schulze et al. 2024). 
  In the panel {\bf a} the model includes the scattering in \feii\ in the ejecta and in the layer of fragmented CDS. 
  The model in the panel {\bf b} considers only the scattering inside ejecta.
  }
	\label{fig:fe2}
\end{figure}

\section{Interpretation of problematical spectral features}

\subsection{\feii\ narrow lines}

High-velocity narrow absorption lines of \feii\ 4924, 5018 и 5169\,\AA\ are seen 
   in SN~2018ibb spectra as long as three months starting the light maximum (Schulze et al. 2024).  
Their origin in the spectrum of SN~1999as (SLSN) Kasen (2004) attributed to the  
  absorption in the CDS between forward and reverse shocks. 
However, he emphasized the drawback of this explanation related to the narrow CDS
 spatial width. 
 
This problem is resolved, if one takes into account that the CDS deceleration brings about 
 a thin shell fragmentation due to the Rayleigh-Taylor instability. 
The fragmentation generates an ensemble of thin two-dimensional fragments of the cold gas mixed with a hot gas of the forward shock, e.g.,  demonstrated by three-dimensional simulations (Blondin \& Ellison 2001). 
 This picture has been already implemented in the model of the SN~2002ic spectrum (Chugai et al. 2004).
Noteworthy, apart from narrow lines there is a broad component of \feii\ that forms in the undisturbed ejecta, so the proposed model for \feii\ lines, actually, is two-component.
 
The transfer of the resonant radiation in a layer composed by an ensemble of random 
 absorbing fragments with sizes substantially smaller than the shell radius 
 ($l \ll r$) can be described in terms of a "macroscopic" Sobolev approximation (Chugai et al. 2004).
The line local optical depth along a direction $\vec{s}$ with cosine $\mu = \cos{\theta}$ of 
the angle relative to the radius is determined by the average number of fragments on 
the resonant Sobolev length $l_S = u_t|dv_s/ds|^{-1}$ ($u_t$ is a turbulent velocity). 
The average number of fragments on the length $l_S$ (or geometric optical depth) is 
$\tau_g = n_f\sigma_fl_S$, where $n_f$ is the number density of fragments and 
$\sigma_f$ is the average fragment cross section. 
Combined with the average line optical depth of a fragment ($\tau_f$) 
 this gives us an "effective" optical depth for the local absorption (scattering) 
\begin{equation}
	\tau = \tau_g[1 - \exp{(-\tau_f)}]\,.
\end{equation}
It is convenient to use the factorization $\tau_g = Q\phi(\mu)$, where $\phi(\mu) = 1$ 
  for the homologous expansion $v = r/t$,  $\phi(\mu) = 1/(1 - \mu^2)$ for the 
  expansion with the constant velocity, and $Q\sim1$, along with $\tau_f$, are free parameters. 
The homologous expansion corresponds to the undisturbed supernova ejecta, whereas 
  the constant velocity describes approximately a downstream flow 
  of the forward shock (Chevalier 1982a).

Seven strong \feii\ lines in the range of 4900-5350\,\AA\ are selected from VALD database
  (Kupka et al. 2000) assuming Boltzmann population at $T = 8000$\,K. 
Of these, three lines belong to the multiplet 42, other fall in the range $\gtrsim 5200$\,\AA\    
  and have excitation potential by only  $\approx 0.3$ eV higher than that of multiplet 42;
   this means that the choice of the excitation temperature is not critical.
    
Model parameters include the photosphere velocity $v_p$, ejecta boundary velocity 
 $v_0$, the expansion velocity $v_{cds}$ of the CDS fragment layer, and the relative width of this 
  layer $\delta = \Delta R/R$. 
The optical depth of the fiducial line 5018\,\AA\ in ejecta is adopted to vary as 
  $\tau = \tau_p(v_p/v)^9$.
The optimal photospheric velocity is found to be $v_p = 8000$\kms, similar to that by 
  Schulze et al (2024).
Due to the steep drop of the optical depth along the velocity, the ejecta component is not 
  sensitive to the boundary velocity that is adopted to be $v_0 = 11500$\kms.
The optical depth of fragments is assumed to be large $\tau_f \gg 1$, which suggests 
 the primary role of the $Q$ value.
 
The spectrum is calculated by the Monte Carlo technique assuming pure scattering in lines.
The quasi-continuum spectrum is adjusted to fit the observed continuum in the wavelength range 
  of interest. 
The shown model spectrum along with the observed one at $t_{max} + 33$ days (Figure \ref{fig:fe2}a) 
  is calculated for $v_{cds} = 10300$\kms, $\delta = 0.22$, and  $Q = 0.45$.
The case without scattering in the layer of fragmented CDS is shown in Figure \ref{fig:fe2}b;  
  parameter variations in this case do not result in the  better fit to the observed spectrum. 

The major outcome of the model comparison is that the inclusion of the scattering in the 
  layer of fragments significantly improves the agreement with the observed spectrum, so 
   the model with scattering in the fragmented CDS is preferred. 
 The value $Q = 0.45$ indicates that the required mixing of the CDS fragments in the forward shock is moderate (cf. Appendix A1).
The scattering in \feii\ lines of ejecta also matters; particularly, it is responsible for the concave shape of the blue part of the 5018\,\AA\  emission maximum.
The optimal value of the optical depth for this line at the photosphere is $\tau_p = 0.87$.

The presented two-component model describes the observed \feii\ spectrum fairly well and 
  thus supports the formation of \feii\ narrow lines in the CDS 
  (Kasen 2004) that, however, necessarily should include the CDS fragmentation and mixing.
 Interestingly, the phenomenon of narrow \feii\ lines of SN~2018ibb  essentially turns out 
 to be an analogue of the 
  high-velocity narrow absorption (HVNA)  \ha\ in SNe~IIP at the stage of 50-70 days (Chugai et al. 2007).

\subsection{[\oi] doublet and [\siii] doublet}

The model for [\oi] doublet affected by [\siii] lines suggests homologously expanding 
 ($v = r/t$) spherical ejecta with emission sources of optically thin 6300, 6364\,\AA\ lines 
  distributed as $\epsilon \propto v^{-k}$. 
The model admits a central spherical zone of a radius $v_h$ devoid of oxygen.
This presumably imitates the PISN one dimensional unmixed model in which oxygen is distributed 
 in external layers, likewise in the model He130 (Kasen et al. 2011). 
In our model the population of the lower level of \siii\ 6347, 6371\,\AA\ doublet 
 (upper level of a relevant ultraviolet transition $n_2$) has the stepwise distribution, 
  $n_2 = const$ for $v \lesssim v_{si}$ and  $n_2 = 0$ for $v > v_{si}$.
The local optical depth has a similar distribution assuming a constant turbulent (thermal) 
  velocity along the radius.
 The line optical depth ratio $\tau(6371)/\tau(6347) = 0.59$ (NIST database).
 The model [\oi] doublet is determined by four parameters: power index $k$, 
  $\tau(6347)$, $v_{si}$, and $v_h$.
 In the case of large optical depth $\tau(6347) \gg 1$ we have three parameters ($k$, $v_{si}$, $v_h$).
 
Apart from the line radiation the model takes into account a quasi-continuum that is 
 represented by a linear function of the wavelength.
The observed quasi-continuum in the range of 6000--6600\,\AA, in fact, is a broad emission 
 band that is composed by the blend of \feii\ emission lines [e.g., SN~2002ic spectrum 
 (Chugai et al. 2004, Figure 10)].
 Quasi-continuum sources in the model are distributed according to the same law as emission 
  sources of [\oi] doublet.
The velocity at the ejecta boundary is determined by the short wavelength boundary of the doublet and is equal to $v_0 = 11500$\kms.

\begin{figure}
	\centering
	\includegraphics[trim=0 100 0 210, width=\columnwidth]{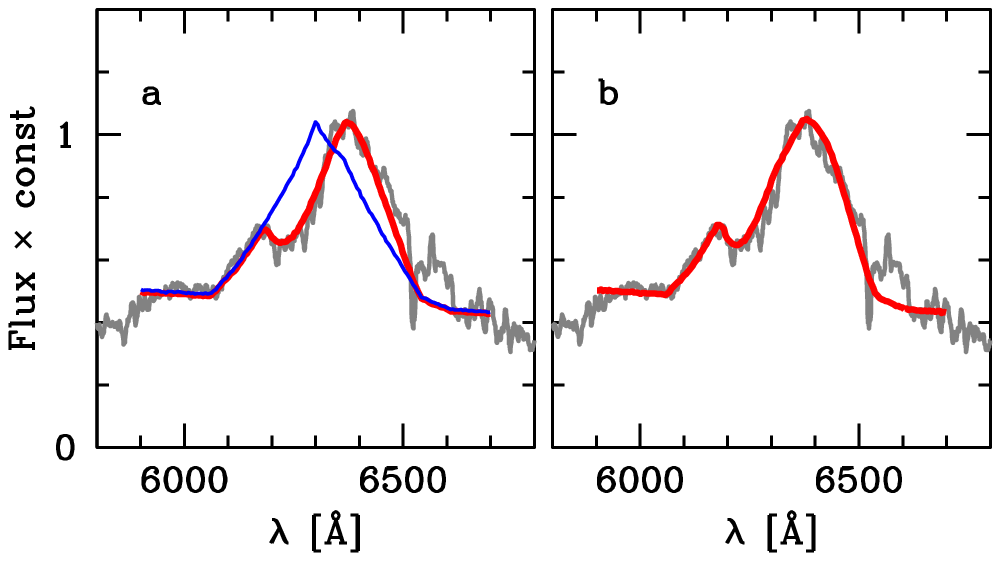}
	\caption{
		The model [\oi] doublet ({\em red} line) with the scattering in \siii\ 6347, 6371\,\AA\ lines compared to the observational spectrum at $t_{max} + 287$ days
		(Schulze et al. 2024). {\em Thin blue} line is the doublet profile in the absence of  scattering in \siii.
		{\em Left} ({\bf a}) the model without central hole in the oxygen distribution, 
		{\em right} is the case with the hole ($v_h = 3000$\kms).	
	}
	\label{fig:o2}
\end{figure}

The [\oi] doublet calculated by the Monte Carlo technique, with the scattering in \siii\ lines, 
   are shown in Figure \ref{fig:o2} for two cases of the oxygen distribution:
   without hole ($v_h = 0$) and with the hole ($v_h = 3000$\kms) in both cases with the boundary velocity  $v_{si} = 7300$\kms\ for the $n_2(\mbox{\siii})$) distribution.
 The uncertainty of $v_{si}$ value is $\pm300$\kms. 
 The power law index of the oxygen distribution is $k = 1.23$ in the first case and 
 $k = 1.65$ in the second case.
Both versions provide acceptable fit, which means that the result is not sensitive to
  the presence of the hole in the oxygen distribution, if  $v_h \lesssim 3000$\kms.
 In the shown models $\tau(6347) = 5$, yet the same result is obtained for $\tau(6347) = 4$.
The estimate in Appendix A2 shows that these $\tau(6347)$  values are realistic.
To summarize, the observed selective absorption in [\oi] doublet is certainly due to the 
  scattering in the \siii\ doublet.  

\subsection{[\oiii]\,5007, 4959\,\AA\ emission site}

The CSM issue closely linked with the problem of 
  the [\oii] and [\oiii] emission lines at the nebular stage.
Two options are conceivable: (i) emission of the CSM ionized by the shock wave radiation 
(Schulze et al. 2024) and (ii) emission from supernova ejecta ionized by the $^{56}$Co 
  radioactive decay.
The first version predicts verifiable features. 
Indeed, for the CSM constant expansion velocity the profile of an optically thin emission 
  line is rectangular that should be noticeable in the observational profile.

\begin{figure}
 	\centering
 	\includegraphics[trim=0 200 0 80,width=\columnwidth]{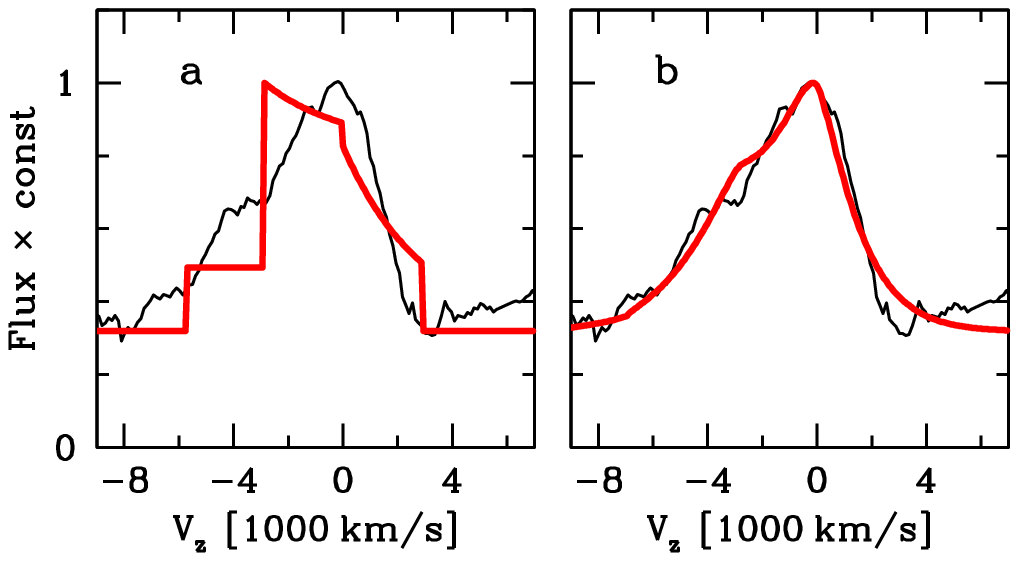}
 	\caption{
 		Model [\oiii]\,5007, 4959\,\AA\ doublet ({\em red} line) originated from 
 		CSM ({\bf a}) and supernova ({\bf b}) compared to the observed spectrum 
 		at the age $t_{max} + 565$ days (Schulze et al. 2024).
 	}
 	\label{fig:oiii}
 \end{figure}

The calculated both versions of [\oiii]\,5007, 4959\,\AA\ emission sites, 
compared to the observed spectrum at $t_{max} + 565$ days, are shown in Figure \ref{fig:oiii}. 
Profile asymmetry especially pronounced at this stage is related to the continuous absorption  
  in the supernova ejecta (Schulze et al. 2024) taken into account in both models. 
In the first option (Figure \ref{fig:oiii}a)
the CS shell is adjusted to the homogeneous absorbing envelope of the radius 
$R_1$ and the continuum optical depth $\tau = 0.5$. 
The CS shell with the expansion velocity of 2900\kms\ and uniformly distributed sources 
  occupies the layer $R_1 < r < R_2 = 1.2R_1$; more extended CS shell does not attain  
   the required occultation.
In the case of the supernova line-emitting site (Figure \ref{fig:oiii}b) the emissivity distribution in the range 
 of $v \leq v_b$ is proportional to the density distribution of the model mod60 (Section 3).  
Absorption sources are uniformly distributed in the same velocity range with the radial optical depth $\tau$.
In the shown optimal case $v_b = 7000$\kms\ and $\tau = 1$.
The shown model profiles suggest that the supernova as the line-emitting site is 
 preferred (see, however, Discussion).

The significant continuous absorption could result from the dust formation 
  in the supernova envelope.
 This is favored by the low effective temperature ($\sim 1000$\,K) at the 
  considered epoch $t \approx 670$ days after the explosion and the high metal abundance   
   (Si, C, Fe, O).
 For the grain radius of 0.1\,mcm the dust amount of $\sim 10^{-4}$\msun\ provides the 
  required optical depth of $\tau \sim 1$.
In fact, the dust mass could exceed the estimated value due to, e.g., a clumpy dust distribution.

\section{Circumstellar envelope}

The circumstellar interaction model is based on the thin shell approximation (Chevalier 1982b; Guiliani 1982).
Apart from the dynamics our model calculates also the luminosity related to the conversion of 
 the X-ray emission from forward and reverse shocks, as well as the diffusion luminosity 
   powered by the radioactive decay (Chugai 2001).
For a given mass and energy, the CS interaction hydrodynamics depends on the supernova density 
 distribution.
 Based on the PISN models (Kasen et al. 2011) I adopt the exponential distribution $\rho\propto \exp{(-v/v_{sc})}$, where 
  $v_{sc}$ depends on the ejecta mass ($M$) and energy ($E$).

\begin{figure}
 	\centering
 	\includegraphics[trim= 0 100 0 160, width=\columnwidth]{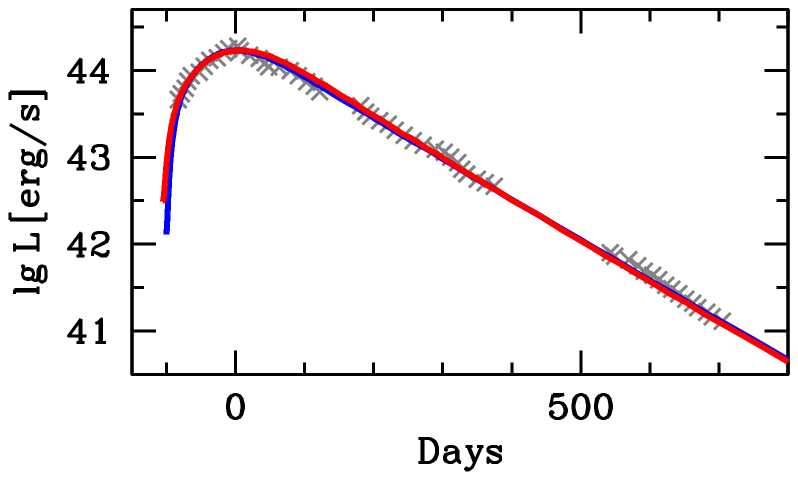}
 	\caption{
 		Bolometric light curve in models mod40 ({\em blue} line) and 
 		mod80 ({\em red}) compared to observational light curve ({\em crosses}). 
 	Both models describe satisfactorily observations and demonstrate $E$ and $M$ parameter 	
 		degeneracy, yet predict similar values of the $^{56}$Ni mass.
 	 	}
 	\label{fig:arn}
 \end{figure}

Mass and energy can be recovered from the light curve modeling without CS interaction based on   
  the Arnett  approximation (Arnett 1980) assuming ejecta uniform density.
Apart from  $M$ and $E$ the model depends on opacity that is adopted to be 0.07\cmsqg\ in line 
  with Schulze et al. (2024).
Another parameter is the mass fraction (0.8) in which the $^{56}$Ni is mixed.
Given degeneracy of $M$ and $E$ parameters two versions are shown: the low mass 
  model (mod40) with 40\msun\ ejecta and the high mass model (mod80) with 80\msun\ ejecta 
  (Figure  \ref{fig:arn}).
The kinetic energy, $^{56}$Ni and the maximum moment ($t_{max}$) for mod40 are equal 
  5 Bethe, 29\msun, 100 days, whereas for mod80 they are 20 Bethe, 32\msun, 105 days, respectively.
Remarkably, despite the large difference of the mass and energy values, the $^{56}$Ni mass differ by only 
 10\% [Arnett rule in action: for compact presupernovae the 
  $^{56}$Ni mass is determined by the bolometric luminosity at the light maximum].

\begin{figure}
	\centering
	\includegraphics[trim=0 140 0 120,width=\columnwidth]{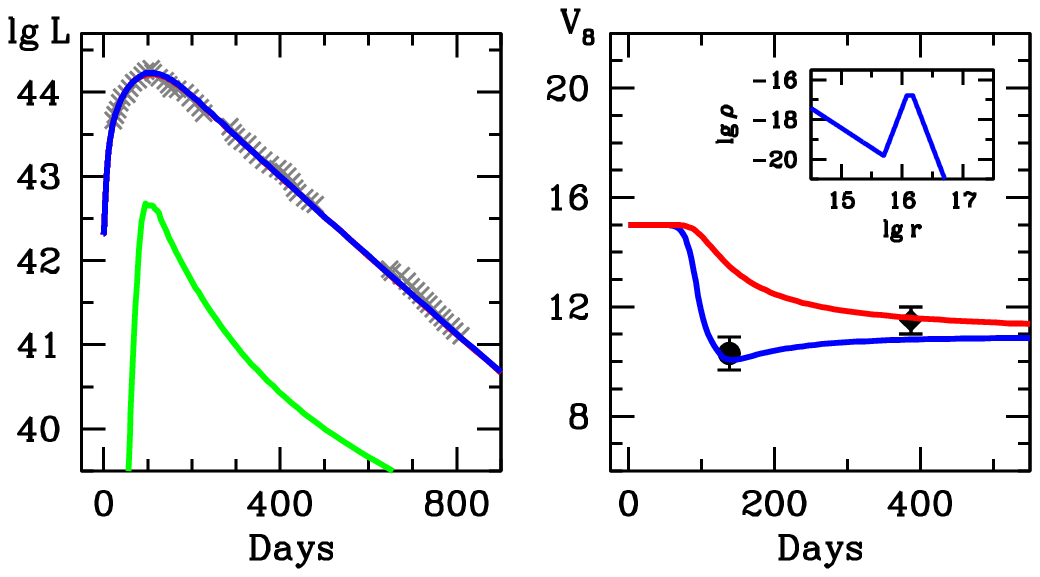}
	\caption{
	Interaction of the supernova in the model mod60 with the CSM. 
	{\em Left.} Bolometric luminosity without the CS interaction ({\em red} line) and the same 
	 with the CS interaction ({\em blue} line) compared to the observational light curve 
	 ({\em crosses}); {\em green} line is the luminosity caused by the conversion of X-rays of the forward and reverse shocks into the optical radiation.
	 The contribution of the interaction luminosity to the total bolometric luminosity 
	 is unnoticeable.  
	{\em Right.} The CDS velocity ({\em blue} line) and boundary velocity of the undisturbed 
	supernova envelope ({\em red} line) compared to the estimates of the CDS  velocity based on \feii\ ({\em circle}) and boundary velocity of the undisturbed supernova envelope 
	from [\oi] ({\em diamond}). {\em Insert} shows density distribution of the CSM along the radius. 
	}
	\label{fig:csi}
\end{figure}

Preliminary computations of the SN/CSM interaction show that the CDS mass of $\approx 0.2$\msun\ required for the 
 narrow \feii\ lines at $t_{max} + 33$ days (cf. Appendix A1) prefers  
  the model mod60 with the mass of 60\msun, the energy of $1.2\times10^{52}$ erg, and $t_{max} = 105$ days (Figure \ref{fig:csi}).
The CSM density is described as $\rho \propto r^{-2}$ with the additional dense shell at the 
  distance of $1.4\times10^{16}$\,cm (Figure \ref{fig:csi}, inset).
This CS density distribution permits one to provide the rapid ejecta deceleration required to 
  describe \feii\ lines.
The interaction model takes into account the CSM expansion velocity of 2900\kms\ (Schulze et   
  al. 2024).
In the shown optimal model (Fgure \ref{fig:csi}) the mass of the CS shell 
  is 0.14\msun\ inside the radius of $3\times10^{16}$\, cm with the density maximum at 
  $1.4\times10^{16}$\,cm.

\section{Discussion}

The paper goal has been to recover the mass and the radius of the CS shell in the vicinity of 
   SLSN SN~2018ibb.
The SN/CSM interaction model combined with velocities of the CDS and supernova 
  ejecta inferred from narrow \feii\ absorption lines and emission [\oi] doublet along with  
  the bolometric light curve, suggest the CS shell mass of 0.14\msun\ inside  $3\times10^{16}$\,cm with the density maximum at $1.4\times10^{16}$\,cm.
  
The expansion velocity of the CSM $v_{cs} = 2900$\kms\ (Schulze et al. 2024), given the 
  CS shell mass, indicates that the ejection of this shell with the moderate energy of 
   $\sim1.2\times10^{49}$\,erg  occured $\Delta t \sim  R_{cs}/v_{cs} \sim 1.5-3$ yr 
   prior the final supernova explosion.  
The absence of helium signature in spectra and the high oxygen velocity (11500\kms)
indicated by [\oi] doublet suggests that SN~2018ibb has been an explosion of C/O core of 
  a supermassive star lost significant fraction of its mass, probably via stellar wind. 
Remarkably, the inferred low mass and low energy of the CS shell 
  essentially eliminates the (would-be) problem of PISN attribution, i.e., the incompatibility between the high energy and $^{56}$Ni mass of SN~2018ibb, on the one hand, and a large CS mass (if this were the case), on the other hand.
   
Despite the formation of narrow \feii\ in the SLSNe has been attributed to 
  the CDS long ago (Kasen 2004), the small CDS thickness left this idea unworkable, 
  until the CDS fragmentation is included in the case of SN~2018ibb.     
Noteworthy, the model for narrow \feii\ in SLSNe essentially is similar to the 
 model for the high-velocity narrow absorption (HVNA) of \ha\ in SNe~IIP spectra (Chugai et al. 2007).
 
The interpretation of the selective absorption in [\oi]  at the nebular stage in terms of 
 the scattering by \siii\ lines permits one to avoid a problem of a large oxygen mass 
   in the CDS required in the alternative conjecture of the absorption in [\oi] doublet lines.
In fact for the realistic CDS mass $<1$\msun\ the absorption in [\oi] lines is 
  negligible.
  
The conjecture on the emission of [\oiii] and [\oii] lines in supernova 
  ejecta seems promissing, yet should be verified in future via the modeling 
  of the oxygen ionization and excitation powered by the $^{56}$Co decay.
On the other hand, the option of the CSM should not be completely rejected.
Indeed, in the case of a clumpy structure of the CSM with low velocity clouds, the line    
  profile  related to the emission of CS clouds shocked, fragmented, and accelerated in the forward shock 
 (Chugai 2018) might be resemblant of [\oiii] and [\oii] lines in SN~2018ibb.
  
The attractive feature of the supernova line-emitting site  is the natural explanation of 
  the [\oiii] doublet blueshift at $t_{max} + 565$ days by the dust formation with the mass  of $\gtrsim 10^{-4}$\msun\  within the  velocity range of $\lesssim 7000$\kms.
The blueshift of [\oiii] 5007, 4959\,\AA\ doublet is seen also at $t_{max} + 287$  day 
  (Schulze et al. 2024), which implies that the dust could be present at this stage as well. 
This possibility is supported by the photometric detection of dust signatures in SLSN 
  SN~2018bsz at $t > t_{max} + 200$ days  (Chen et al. 2021).
   
Noteworthy, in SN~2018ibb dust signature is revealed due to the blueshift  of emission line.
Originally, the blueshift of emission line related to the dust formation was discovered 
 in nova DQ He 1934 (Payne-Gaposchkin \& Wipple 1939),  although without invoking the dust.
For supernovae, the occultation of a line-emitting zone by the dust is demonstrated by the 
  blueshift of [\oi]\,6300\,\AA\ emission in SN~1987A (Lucy et al. 1991).
 
\section{Conclusion}

I conclude with results in the form of short statements.
\begin{itemize}
	\item 
	Modeling the SN~2018ibb interaction with the CSM combined with observed effects indicates the presence of the CS shell with the mass of $\sim 0.14$\msun\ inside the radius 
	 of $\lesssim3\times10^{16}$\,cm and density maximum at $\sim 1.4\times10^{16}$\,cm.
	\item
	Narrow \feii\ absorption lines are reproduced in the model of the radiation scattering by 
	 CDS fragments mixed in the forward shock.
	\item
	 The [\oi] doublet with the unusual effect of a selective absorption is reproduced in the 
	  model that includes scattering in \siii\ doublet; the absorption by [\oi]
	 doublet lines is negligible.
	\item
    I propose the emission of [\oiii] 5007, 4959\,\AA\ doublet in the supernova ejecta; 
     the doublet blueshift in this case is explained by the dust formation in the supernova 
     envelope.

\end{itemize}

\bigskip   
\section{Acknowledgments}

I thank Daniel Kasen for sharing his dissertation text and I thank Steve Schulze for the 
 sent late-time  SN~2018ibb spectrum.

\clearpage

\section{References}

\noindent
Arnett W. D., Astrophys. J., {\bf 237}, 541 (1980)\\
\medskip
Arnett W. D., Astrophys. J.,  Astrophys. J. {\bf 253}, 785 (1982)\\
\medskip
Blondin J. M., Ellison D. C., Astrophys. J. {\bf 560}, 244 (2001)\\
\medskip
Chen T. W., Brennan S. J., Wesson R., et al., ArXiv e-prints
[arXiv:2109.07942] (2021)\\
\medskip
Chevalier R. A., Blondin J. M., Astrophys. J. {\bf 444}, 312 (1995)\\ 
\medskip
Chevalier R. A., Astrophys. J. {\bf 258}, 790 (1982a)\\ 
\medskip
Chevalier R. A., Astrophys. J. {\bf 259}, 302 (1982b)\\
\medskip
Chugai N. N., Mon. Not. R. Astron. Soc. {\bf 481}, 3643 (2018)\\
\medskip
Chugai N. N., Chevalier R. A., Utrobin V. P., Astrophys. J. {\bf 662}, 1136 (2007)\\
\medskip
Chugai N. N., Chevalier R. A., Lundqvist P., Mon. Not. R. Astron. Soc. {\bf 355}, 627 (2004)\\
\medskip
Chugai N. N., Mon. Not. R. Astron. Soc. {\bf 326}, 1448 (2001)\\
\medskip
Giuliani J. L.,  Astrophys. J. {\bf 256}, 624 (1982)\\
\medskip
Heger A., Woosley S. E., Astrophys. J. {\bf 567}, 5326 (2002)\\
\medskip
Kasen D. N., Ph.D. dissertation, University of California, Berkeley (2004)\\
\medskip
Kasen D., Woosley S. E., Heger A., Astrophys. J. {\bf 734}, 102 (2011)\\
\medskip
Kupka F. G., Ryabchikova T. A., Piskunov N. E., Baltic Astronomy {\bf 9}, 590 (2000)\\
\medskip
Lucy L. B., Danziger I. J., Gouiffes C., Bouchet P., 
{\em Supernovae. The Tenth Santa Cruz Workshop in Astronomy and Astrophysics.}
Ed. S.E. Woosley. New York: Springer-Verlag, 1991\\
\medskip 
Moriya T. J., Mazzali P. A., Tanaka M., Mon. Not. R. Astron. Soc. {\bf 484}, 
3443 (2019)\\
\medskip 
Payne-Gaposchkin C., Whipple F. L., Harvard College Obs. Circular {\bf 433}, 1 (1939)\\
\medskip
Wiese W. L., Smith M. W., Miles B. M., {\em Atomic Transition Probabilities, Vol. 2.} Washington, D.C.: Institute for Basic Standards, 1969\\
\medskip
Woosley S. E., Heger A., Weaver T. A., Review Mod. Phys. {\bf 74}, 1015 (2002)\\
\medskip
Schulze S., Fransson C., Kozyreva A., et. al.), Astron. Astrophys. {\bf 683}, A223 (2024)\\

\clearpage

\vspace{1cm}
\centerline{Appendix A1}
\vspace{0.5cm}
\centerline{\large{Parameters $Q$ and $\tau$ for \feii}}

\vspace{0.5cm}
The parameter of the geometric optical depth of the spherical layer that is filled by 
 a chaotic ensemble of flat CDS fragments, produced by the Rayleigh-Taylor instability, 
 can be expressed via the ratio of the total surface area (of one side) fragments $S$ to 
 the area of the spherical CDS $S_0 = 4\pi R^2$, the relative thickness of the layer 
 $\delta = \Delta R/R$, and the ratio of the turbulent velocity to the layer velocity $u_t/v$ 
  (Chugai et al. 2004)
\begin{equation}
Q \approx \left(\frac{S}{S_0}\right)\left(\frac{u_t}{v}\right)\delta^{-1}\,.
\end{equation}
For $\delta = 0.22$ and the ratio $u_t/v \sim 0.1$ (Chevalier \& Blondin 1995)  
    the moderate fragmentation and mixing $S/S_0 \sim 1$ is sufficient for the 
     parameter $Q \sim 0.5$ that is needed for narrow \feii\ lines at $t_{max} + 33$ days. 
  Noteworthy that the large value $S/S_0 \sim 10 - 100$ in the case of SN~2002ic
   (Chugai et al. 2004) is related to the adopted low turbulent velocity.

The optical depth of a typical fragment in the 5018\,\AA\ line for the relatively low 
  mixing degree should be of the order of the CDS optical depth.
On day 133 after the explosion the CDS mass of 0.2\msun\ (model mod60) suggests 
  the barion column density $N_b = 8\times10^{22}$\cmmsq.
For the excitation temperature of 8000\,K, the Doppler width of 10\kms,  and Fe solar abundance 
 $1.3\times10^{-3}$ (by mass) one obtains $\tau(5018\mbox{\AA}) \sim 40$. 
 For the metallicity 1/4 of the solar one (Schulze et al. 2024) the expected fragment optical depth in 
 5018\,\AA\ line is $\sim 10$.
 
\vspace{1cm}

\vspace{1cm}
\centerline{Appendix A2}
\vspace{0.5cm}
\centerline{\large{Optical depth in \siii\ 6347\,\AA} line}

\vspace{0.5cm}

The bolometric luminosity at 390 days after the explosion is $L = 6\times10^{42}$\ergs\
  (Schulze et al. 2024).
This luminosity is equal to the power of the $^{56}$Co decay, which implies the deposition rate 
  in the unit of volume $V$ with the boundary velocity of 7300\kms\ of 
 $\epsilon \approx L/V = 0.96\times10^{-7}$\ergcms. 
The deposited energy is spent on the ionization, excitation. and Coulomb heating with the fraction 
 for the each channel approximately $\psi_1 \approx 1/3$.
The Si mass abundance in the unmixed PISN model He130 (Kasen et al. 2011) in the inner and  
  outer zone of the expanding ejecta is as high as 0.5.   
One can assume that in the mixed case with three major components in the proportion  Fe/Si/O = 2:1:1 the Si abundance is close to 1/4.
This means that the deposition fraction spent on Si is $\psi_2 \approx $ 0.25--0.5.
The excitation potential of the lower level ($^2\mbox{S}_{1/2}$) for 
  \siii\ 6347\,\AA\ line is $E_{12} = 8.12$ eV.
The corresponding transition (UV2) is among six ultraviolet transitions from the ground  
  level with the branching ratio of $\psi_3 = 0.05$ (Wiese et al. 1969).    
As a result, the efficiency of the non-thermal excitation of the lower level of 
  \siii\ 6347\,\AA\ ($^2\mbox{S}_{1/2}$) is $\eta = \psi_1\psi_2\psi_3 \approx (4 - 8)\times10^{-3}$.

The excitation degree ($n_2/n_1$) can be determined from the balance equation  
\begin{equation}
  n_2A_{21}\beta_{12} = \eta \epsilon E_{12}^{-1}\,,	
\end{equation}
where $\beta_{12} \approx 1/\tau_{12}$ is the local escape probability for the resonant photon.
With the above numerical values one obtains $n_2/n_1 = (1-2)\times10^{-7}$.
The density at the velocity 7300\kms\ in the model mod60 is $\rho = 2\times10^{-16}$\gcmq, 
 in which case the Si number density is $n(\mbox{Si}) = 4\times10^6$\cmq. 
Assuming that all the Si is singly ionized the found excitation degree results in 
 the optical depth $\tau(6347) \sim 15-30$ at the level 7300\kms\ at the observational epoch ($t_{max} + 287$ days).

\end{document}